\documentclass{article}
\usepackage{ijcai21}

\usepackage{times}
\usepackage{soul}
\usepackage{url}
\usepackage[hidelinks]{hyperref}
\usepackage[utf8]{inputenc}
\usepackage[small]{caption}
\usepackage{graphicx}
\usepackage{amsmath}
\usepackage{amsthm}
\usepackage{booktabs}
\usepackage{algorithm}
\usepackage{algorithmic}
\usepackage{balance} 
\usepackage[english]{babel} 
\usepackage{listings} 
\usepackage{color}
\graphicspath{ {C:/Users/kdmus/Desktop/IJCAI/IJCAI/images/} }
\usepackage{cancel}
\usepackage{pgfplots}
    \pgfplotsset{
        table/search path={ {C:/Users/kdmus/Desktop/IJCAI/IJCAI/images/} },
    }

\newcommand{\BibTeX}{\rm B\kern-.05em{\sc i\kern-.025em b}\kern-.08em\TeX}
\newcommand\aug{\fboxsep=-\fboxrule\!\!\!\fbox{\strut}\!\!\!}

\title{Learning to Win, Lose and Cooperate through Reward Signal Evolution}

\author{
Rafa\l{} Muszy\'nski$^1$
\and
Katja Hofmann$^2$\and
Jun Wang$^{1}$
\affiliations
$^1$University College London\\
$^2$Microsoft Research Cambridge, UK
\emails
\{r.muszynski, jun.wang\}@cs.ucl.ac.uk,
katja.hofmann@microsoft.com
}

\begin{document}

\maketitle

\begin{abstract}
Solving a reinforcement learning problem typically involves correctly prespecifying the reward signal from which the algorithm learns. Here, we approach the problem of reward signal design by using an evolutionary approach to perform a search on the space of all possible reward signals. We introduce a general framework for optimizing $N$ goals given $n$ reward signals. Through experiments we demonstrate that such an approach allows agents to learn high-level goals - such as winning, losing and cooperating - from scratch without prespecified reward signals in the game of Pong. Some of the solutions found by the algorithm are surprising, in the sense that they would probably not have been chosen by a person trying to hand-code a given behaviour through a specific reward signal. Furthermore, it seems that the proposed approach may also benefit from higher stability of the training performance when compared with the typical score-based reward signals.
\end{abstract}

\section{Introduction}

Typically, the most important goal of an agent in a reinforcement learning (RL) setting is to maximize the cumulative reward function over the long run, based on the immediate reward it receives at each timestep from the environment. Consequently, the reward is the main source of information for the algorithm allowing it to decide whether a given interaction within its environment was positive or not. In a way, the reward plays a similar role for an agent, as the biological sensations do for humans ~\cite{sutton2018reinforcement}. 

Oftentimes, for each RL problem there exists some prespecified goal function to maximize in designer's mind. Generally, in simple environments, it is straightforward to translate the desired intention (our goal) into the reward signal. The process of designing a reward signal usually comes down to trying many variants of the settings, and observing the process of algorithm's improvement on the goal along the way \cite{sutton2018reinforcement}. Yet, one might argue that as we move towards more difficult environments, especially those that modern deep reinforcement learning (DRL) tries to study, this process can quickly become unpredictable and lead to surprising results \cite{muszynski2017happiness}. Additionally, RL algorithms have also been shown to be able to ``hack" their own goal functions to deliver the reward in unintended and unexpected ways \cite{amodei2016concrete}.

As a result, in this work we concentrate on exploring the space of all possible reward signals in a RL setting in order to bypass hand-engineering of the reward signal. We propose an algorithm for evolving the reward signal at each timestep of agents' interaction with the environment in order to optimize $N$ goals that are of interest to the designer of the system. We demonstrate that the algorithm can work in the domain of the game of Pong when trying to learn the goals of winning, losing and cooperating.

We argue that a number of the goal and reward signal pairs found by the algorithm are non-obvious and probably would not have been specified a priori. We also find differences in the training performance between our approach and typical score-based reward signals hand-coded in Pong. Concretely, for some of the goals the evolutionary approach yielded better stability of training or an overall better performance on a given goal.

\section{Background} \label{sec:background}

Reinforcement learning (RL) is an area of machine learning that describes the learner as an agent in an environment \cite{sutton2018reinforcement}.
Mathematically, the setting is usually formalized as a \emph{Markov decision process (MDP)}.

The goal of the agent, the goal of learning in an MDP, is to find a policy $\pi$ that maximizes the cumulative reward by best mapping states to action selection probabilities. The cumulative reward is described by the sum of immediate scalar reward $r$ over $t$ timesteps: $\sum_{1}^{t} r_t$.

In the case when the model of the environment is not available, RL can be used to find $\pi$. \emph{Q-learning} \cite{dayan1992q} is arguably the most famous example of an RL algorithm. It is a model-free temporal difference algorithm. In Q-learning, a value function $Q(s,a)$ is calculated over state-action pairs. It is updated based on the immediate reward and the discounted expected future reward in the following way:
\begin{align}
Q(s_t, a_t) \gets & Q(s_t, a_t) + \alpha [r_{t+1} + \nonumber \\ &+ \gamma \max_{a} Q(s_{t+1}, a) - Q(s_t, a_t)]
\end{align}

In the update rule above, $\gamma$ is the discount factor for future rewards, and $\alpha \in [0,1]$ is the learning rate that determines how quickly $Q$ is updated based on new reward information.
Q-learning is proven to converge to the optimal policy $\pi^*$, given “sufficient” number of updates for each state-action pair, and a decreasing learning rate $\alpha$.

The growing amount of data and much better technology to work with it, have both contributed greatly to the recent reemergence of interest in neural networks. As a result, neural networks have grown from ``shallow" models of a few layers to ``deep" models that have quickly shown impressive practical power in the realm of image recognition and beyond \cite{lecun2015deep,schmidhuber2015deep}.

Moreover, it turned out that in complex environments (large number of states) it is possible to approximate the action-value function by a deep neural network with parameters $\theta$. This approach is known as a Deep Q-learning algorithm (DQN) \cite{mnih2015} - and has given birth to Deep Reinforcement Learning (DRL). Concretely, in DQN $Q_\theta^*$ is obtained by minimizing the loss function:
\begin{equation}
L_{t}(\theta) = E_{h_{t},a_{t},r_{t},h_{t+1}}[(y_t - Q_\theta^\pi(h_t, a_t))^2],
\end{equation}
where experience reply is a tuple $h_t = (s_t, a_t, s_{t+1}, r_{t+1})$, $y_t = r_t + \gamma \max_{a'}Q_\theta^\pi(a_{t+1}, h_{t+1})$,
and where $\pi$ is an $\epsilon$-greedy policy that takes an action $argmax_{a_t}Q^\pi(a_{t}, h_{t})$ with probability $1 - \epsilon$, or
takes a random action otherwise.

The methods presented above focus on estimating value functions in order to solve RL problems. The problem can also be approached with evolutionary computation \cite{sutton2018reinforcement,eiben2015introduction}. Its main building block is an \emph{evolutionary algorithm}, the main components of which are: representation (definition of individuals), evaluation function (or fitness function), population, parent selection mechanism, variation operators, recombination and mutation, survivor selection mechanism (replacement) \cite{eiben2015introduction}.

In this work, we concentrate on using DQN as an approach to learn each goal based on the rewards that are generated by using a subset of elements of the evolutionary algorithm (see Algorithm \ref{alg:re}).

\section{Related work} \label{sec:relWork}

We present the literature connected to our work by looking through the following lenses: approaches to learning a reward signal specifically through evolution and evolutionary methods with the focus on deep reinforcement learning.

\subsubsection{Reward evolution}

The most closely related works to that presented in this paper are those concentrating on the evolution of the reward. The observation leading to this approach can be the fact, that the reward can be seen as one of the variables in the general RL setting. As a result, it can be optimized, and evolutionary algorithms are often a good choice for any optimization task. The proposed solution can then be evaluated against the designer's goal, and the worst performing solutions eliminated, leading to an improved result.

To the best of our knowledge, the work by Singh et al. [2009] is the closest in spirit to ours. The work places the design of a reward function in an evolutionary context. The paper formulates an optimal reward function given a fitness function and some distribution of environments. It then uses exhaustive search over the parameter space of the reward function. In the experiment, the authors find that the reward found through this process can differ from the one intended by the fitness function, but can still be advantageous \cite{singh2009rewards}.

Our work differs in the sense that we present the evolution of the reward in a complex domain of a Pong game, in an environment with an opponent present. Whereas Singh et al. [2009] concentrate on experiments in a setting with predefined features to define the combinatorial space of reward functions, we consider the mutations of the reward at each timestep. While Singh et al.[2009] present an emergence of reward functions, our approach additionally shows an emergence of complex behaviours of an agent.

Singh et al. [2009] extended their work with the focus on the problem of intrinsic versus extrinsic motivation \cite{singh2010intrinsically}. More recently, the framework presented by Singh et al. [2009] has been expanded to the space of reward functions spanned by a given set of feature states and multi-objective evaluation function \cite{grunitzki2017flexible}.

Genetic Programming has also been used in order to evolve a population of reward functions \cite{niekum2010evolved,niekum2011evolution,niekum2010genetic}. In order to search for reward functions, the search is preformed over the space of programmes generating the rewards. The approach is proposed as a method to ``alleviate the difficulty of scaling reward function search" and replace the exhaustive search used by Singh et al. [2009].

Another approach has been to use evolutionary computation to evolve a population of agents' behaviour, allowing them to inherit the rewards as a secondary optimization process \cite{ackley1991interactions}. The rewards are mutated, hence the fitness can increase over time. There are two neural networks involved. The first evaluation network is fixed and uses inherited weights to produce a scalar reward. The second - action network - is also inherited, but trainable after initialization.

\subsubsection{Evolutionary methods in deep reinforcement learning} 

Lastly, we broadly describe the context of using evolutionary methods specifically in a deep reinforcement learning setting.

Perhaps the first study to present the applicability of evolutionary methods in the context of complex environments was by Salimans et al. \cite{salimans2017evolution}. It showed that using evolutionary strategies and enough computational power was enough to learn to perform well in the domain of Atari games. The research mixing the worlds of DRL and evolution has quickly followed. Ideas from evolutionary computation have been mostly applied to DRL by creating populations of agents which undergo mutations and increase the diversity of strategies created for the agents. Then, when agents interact, the process of elimination guarantees high quality of the resulting solution. Such an approach was part of an algorithm mastering a difficult domain of Starcraft II \cite{alphastarblog}, and the evolutionary aspects of that system have also been investigated \cite{arulkumaran2019alphastar}. Another study drawing on ideas from evolution in the DRL setting is Jaderberg et al. \cite{jaderberg2018human} with an objective of ``the emergence of complex cooperative agents." The goal is achieved by applying a variant of population-based training \cite{jaderberg2017population} to DRL agents.

Recently, the reward search has been automated by adding an additional, evolutionary layer over standard RL layer, to find the reward that maximizes the task objective (through proxy rewards \cite{chiang2018learning} - parameterized versions of the standard environment rewards) in a continuous control task \cite{faust2019evolving}.

To the best of our knowledge, ours is the first study to concentrate specifically on the problem of learning a reward signal in a complex, multi-agent setting through an evolutionary algorithm without any prespecified features used as combinations to mutate the reward function. We create a population of agents by training them separately on different mutations of the reward signal. The reward signal is not prespecified, but its evolution allows us to achieve a set of complex and diverse behaviours.

\begin{algorithm}[tbp]
	\caption{Reward signal evolution algorithm}
	\label{alg:re}
\begin{algorithmic}[1]
	\STATE initialize learning algorithm $A$
	\STATE initialize $i$ goal functions $G_i(r)$
	\STATE initialize $j$ unique random reward sequences $r_j$
	\STATE store $\textbf{r} = \{i \Vert r_j\}_{i{\times}j}$
	\STATE burn-in: train $A$ on $r_j$ for $M$ timesteps
	\REPEAT
	\STATE train $A(r_j)$ until $M$ timesteps
	\STATE test $A(r_j)$ for $m$ timesteps
	\STATE calculate $G_i(r_j)$ on test data
	\FOR{$i$}
	\STATE remove $ argmin \bar G_i(r_j)$ from $\textbf{r}$
	\STATE add $p$ unique random mutations of $r_j$ to $\textbf{r}$
	\ENDFOR
	\STATE $M := M + M$
	\UNTIL{$convergence$}
\end{algorithmic}
\end{algorithm}

\section{Reward signal evolution}

In this section we concentrate on the details of our approach. If we think about the interplay between the number of reward signals and goal functions, we can distinguish the following three cases: a) $1$ to $1$; b) $n$ to $1$; c) $n$ to $N$.

Usually, in RL we have a situation in which there is one predefined reward signal, designed with one assumed goal in mind - we call it the $1$ to $1$ case. One can also have a situation, when many reward signals ($n$) can be explored to find the one best optimizing our goal ($n$ to $1$). The last combination is the one in which we have $n$ reward signal candidates, and use them to optimize for many different goals $N$ ($n$ to $N$). The Algorithm \ref{alg:re} presented in this section can be used to cover these cases, and can be seen as a general framework of sorts.

Concretely, the proposed approach (see Algorithm \ref{alg:re}) requires us to choose our learning algorithm, $1$ or many goal functions we are interested in (can be seen as fitness functions), and $1$ or many unique random reward sequences (here, the goal is a function representing our intended behaviour of the agent, whereas the reward is the instantaneous scalar quantity received by the agent from the environment at each timestep). The algorithm of choice is then trained on the initial reward signals, and its performance on the specified goal functions is checked periodically. At that point two things happen - first, we eliminate the rewards leading to the worst result for each of the goals, and second, we add the required number of new reward signal mutations to the population. In the example provided in Section \ref{sec:Experiment} the convergence means that we have eliminated all the possible reward signal mutations, and are left with one winner per each goal. In general, in many examples one could create almost infinitely many reward signals. That could be seen as the \emph{mutation} phase. Once you do not mutate any new reward signals, the algorithm enters the \emph{elimination} phase, until we are left with top $k$ solutions that we require.

Moreover, we distinguish the following ways to choose $r_j$ in Algorithm \ref{alg:re}:

\begin{itemize}
  \item \textbf{Fixed} - the typical way for setting a reward signal in RL
  \item \textbf{Semi-evolutionary} - the algorithm is free to mutate the reward for the set of rules specified by the expert (e.g. position of the ball in a pong game, or for interacting with specific elements of the environment)
  \item \textbf{Evolutionary} - the algorithm is free to mutate the reward at any timestep $t$.
\end{itemize}

Finally, we note that the evolutionary approach for choosing $r_j$ can be seen as a \emph{reward sampling scheme}, sampling from the distribution of all possible reward signals.

\section{Experiment}\label{sec:Experiment}

\subsection{Set-up}

\begin{figure}[tbp]
	\centering
	\includegraphics[width=0.2\textwidth]{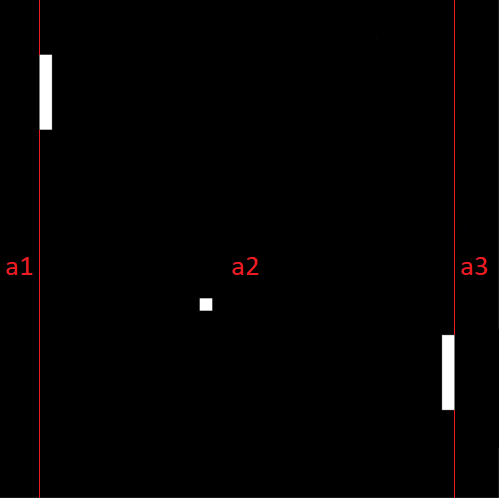}
	\caption{The Pong game screen showing three rectangular regions (a1, a2 and a3 separated by vertical red lines) in which the reward signal can be mutated (it can take either a value of 0 or 1). }
	\label{fig:pong_regions}
\end{figure}

To illustrate how Algorithm \ref{alg:re} could work in practice, we chose the Pong game environment, as shown in Figure \ref{fig:pong_regions}.

In all experiments we concentrate on controlling the paddle on the right side of the screen.
We use DQN \cite{mnih2015} as the base algorithm $A$ in Algorithm \ref{alg:re}.

Typically, RL algorithms have one reward function to optimize for. Here, we give the algorithm three reward functions ($i = 3$). Concretely, the goals are:
\begin{itemize}
  \item \textbf{Winning}, maximizing the cumulative number of points scored by the agent ($G_1(r)$)
  \item \textbf{Losing}, maximizing the cumulative number of points lost by the agent ($G_2(r)$)
  \item \textbf{Cooperation}, maximizing the time spent playing until the point is lost by either of the players ($G_3(r)$).
\end{itemize}

We use the semi-evolutionary way to choose $r_j$ by introducing \emph{mutation regions} to lower the computational cost of the algorithm and have a better intuition and control of the obtained results.
To generate the reward sequences, $r_j$, the algorithm can choose from the values of either 0 or 1, in any of the reward regions $a1$, $a2$, $a3$ depicted in Figure \ref{fig:pong_regions}. That means that there is one unique reward signal when the ball passes either of the paddles, and another one when the ball is between the paddles.
The three regions, combined with a 0 or 1 reward signal in each of them, yield a total of 8 unique reward signals that the algorithm can discover through mutations. We set the initial value of $j$ to 3. Furthermore, we initialize the parameters $M$ and $m$ to have the values of 1 million and 100,000 respectively. That means that we perform the elimination and mutation step of the algorithm after each 1 million timesteps, based on the test results from further 100,000 timesteps. Lastly, after each elimination, we add two new unique reward signals to the population ($p = 2$).

Our aim in choosing this experiment was to first, verify if the proposed algorithm would work at all. Second, limiting the number of possible mutations allows us to perform an exhaustive search on the space of the reward signals and thus have better comparison with the typical score-based approach baseline. We also acknowledge the complexity and scalability issues of the proposed approach, which is typical for evolutionary approaches, however those were not the main focus of our work at this stage.

\subsection{Results}

First, we randomly generated three reward signals, and obtained the following values of $r_j$: $r_1 = 000$, $r_2 = 001$, and $r_3 = 011$. As an example, $r_2 = 001$ corresponds to the reward signal of 1 when the ball passes the paddle on the right, and 0 otherwise (see Figure \ref{fig:pong_regions}). The DQN algorithm was trained on each of the reward signals $r_j$ for 1 million timesteps, as a burn-in phase. Then, the algorithm entered its main loop. After the next 1 million timesteps of training, each $A(r_j)$ was tested for 100,000 timesteps. The values of $G_i(r_j)$ were calculated, and the reward signals corresponding to the lowest value for each goal were eliminated from $\textbf{r}$. Lastly, two unique reward signal mutations were added: $r_4 = 101$ and $r_5 = 110$. Hence, the state of $\textbf{r}$ after the first iteration was given by:

\begin{equation}
\textbf{r} =
\begin{Bmatrix} 
1 \Vert \cancel{000} & 001 & 011 &\aug& 101 & 110\\ 
2 \Vert 000 & 001 & \cancel{011} &\aug& 101 & 110\\ 
3 \Vert \cancel{000} & 001 & 011 &\aug& 101 & 110 
\end{Bmatrix}
\end{equation}

The algorithm continued to run for six more iterations. We present the steps of that process in a graphical form in Figures 2-4.
The figures capture the creation of each new mutation of the reward signal, its score for a given goal, and the reward signals eliminated along the way. Ties (in case when different reward signals led to the same outcome for a given goal) were decided at random.

Additionally, we present each of the iterations in a matrix form, to show the outcome of each loop in more detail:

\begin{equation}
\begin{split}
\textbf{r} = &
\begin{Bmatrix} 
1 \Vert \cancel{000} & 001 & 011 &\aug& 101 & 110\\ 
2 \Vert 000 & 001 & \cancel{011} &\aug& 101 & 110\\ 
3 \Vert \cancel{000} & 001 & 011 &\aug& 101 & 110 
\end{Bmatrix}
\\
\xrightarrow{3M}&
\begin{Bmatrix} 
1 \Vert \cancel{001} & 011 & 101 & 110 &\aug& 010 & 100\\ 
2 \Vert 000 & 001 & 101 & \cancel{110} &\aug& 010 & 100\\ 
3 \Vert \cancel{001} & 011 & 101 & 110 &\aug& 010 & 100
\end{Bmatrix}
\\
\xrightarrow{4M}&
\begin{Bmatrix} 
1 \Vert 011 & \cancel{101} & 110 & 010 & 100 &\aug& 111\\ 
2 \Vert 000 & 001 & 101 & 010 & \cancel{100} &\aug& 111\\ 
3 \Vert 011 & \cancel{101} & 110 & 010 & 100 &\aug& 111
\end{Bmatrix}
\\
\xrightarrow{5M}&
\begin{Bmatrix} 
1 \Vert 011 & 110 & 010 & 100 & \cancel{111}\\ 
2 \Vert 000 & 001 & 101 & \cancel{010} & 111\\ 
3 \Vert 011 & 110 & 010 & 100 & \cancel{111}
\end{Bmatrix}
\\
\xrightarrow{6M}&
\begin{Bmatrix} 
1 \Vert \cancel{011} & 110 & 010 & 100\\ 
2 \Vert 000 & 001 & 101 & \cancel{111}\\ 
3 \Vert \cancel{011} & 110 & 010 & 100
\end{Bmatrix}
\\
\xrightarrow{7M}&
\begin{Bmatrix} 
1 \Vert 110 & \cancel{010} & 100\\ 
2 \Vert 000 & \cancel{001} & 101\\ 
3 \Vert 110 & 010 & \cancel{100}
\end{Bmatrix}
\\
\xrightarrow{8M}&
\begin{Bmatrix} 
1 \Vert \cancel{110} & 100\\ 
2 \Vert 000 & \cancel{101}\\ 
3 \Vert 110 & \cancel{010}
\end{Bmatrix}
\end{split}
\end{equation}

Eventually, the surviving reward signal combinations for each of the goals were:
\begin{itemize}
  \item \textbf{Winning}: $r_7 = 100$
  \item \textbf{Losing}: $r_1 = 000$
  \item \textbf{Cooperation}: $r_5 = 110$
\end{itemize}

\begin{figure}[t]
\centering

\scalebox{0.55}{
\begin{tikzpicture}
    \begin{axis}[xlabel=timesteps (in million), ylabel=won, legend pos=outer north east]
        \addplot[blue, mark=x] table[x=M, y=000] {win.dat};
        \addlegendentry{000}
        \addplot[red, mark=o] table[x=M, y=001] {win.dat};
        \addlegendentry{001}
        \addplot[green, mark=*] table[x=M,y=010] {win.dat};
        \addlegendentry{010}
		\addplot[cyan, mark=-] table[x=M,y=100] {win.dat};
        \addlegendentry{100}
        \addplot[violet, mark=|] table[x=M,y=011] {win.dat};
        \addlegendentry{011}
        \addplot[magenta, mark=triangle] table[x=M,y=101] {win.dat};
        \addlegendentry{101}
        \addplot[gray, mark=+] table[x=M,y=110] {win.dat};
        \addlegendentry{110}
        \addplot[orange, mark=oplus] table[x=M,y=111] {win.dat};
        \addlegendentry{111}
    \end{axis}
\end{tikzpicture}
}
\caption{Survival and fitness of a given reward signal when the goal is \textbf{winning}.}

\bigskip

\scalebox{0.55}{
\begin{tikzpicture}
    \begin{axis}[xlabel=timesteps (in million), ylabel=lost, legend pos=outer north east]
        \addplot[blue, mark=x] table[x=M, y=000] {lose.dat};
        \addlegendentry{000}
        \addplot[red, mark=o] table[x=M, y=001] {lose.dat};
        \addlegendentry{001}
        \addplot[green, mark=*] table[x=M,y=010] {lose.dat};
        \addlegendentry{010}
		\addplot[cyan, mark=-] table[x=M,y=100] {lose.dat};
        \addlegendentry{100}
        \addplot[violet, mark=|] table[x=M,y=011] {lose.dat};
        \addlegendentry{011}
        \addplot[magenta, mark=triangle] table[x=M,y=101] {lose.dat};
        \addlegendentry{101}
        \addplot[gray, mark=+] table[x=M,y=110] {lose.dat};
        \addlegendentry{110}
        \addplot[orange, mark=oplus] table[x=M,y=111] {lose.dat};
        \addlegendentry{111}
    \end{axis}
\end{tikzpicture}
}
\caption{Survival and fitness of a given reward signal when the goal is \textbf{losing}.}

\bigskip

\scalebox{0.55}{
\begin{tikzpicture}
    \begin{axis}[xlabel=timesteps (in million), ylabel=co-op, legend pos=outer north east]
        \addplot[blue, mark=x] table[x=M, y=000] {coop.dat};
        \addlegendentry{000}
        \addplot[red, mark=o] table[x=M, y=001] {coop.dat};
        \addlegendentry{001}
        \addplot[green, mark=*] table[x=M,y=010] {coop.dat};
        \addlegendentry{010}
		\addplot[cyan, mark=-] table[x=M,y=100] {coop.dat};
        \addlegendentry{100}
        \addplot[violet, mark=|] table[x=M,y=011] {coop.dat};
        \addlegendentry{011}
        \addplot[magenta, mark=triangle] table[x=M,y=101] {coop.dat};
        \addlegendentry{101}
        \addplot[gray, mark=+] table[x=M,y=110] {coop.dat};
        \addlegendentry{110}
        \addplot[orange, mark=oplus] table[x=M,y=111] {coop.dat};
        \addlegendentry{111}
    \end{axis}
\end{tikzpicture}
}
\caption{Survival and fitness of a given reward signal when the goal is \textbf{cooperation}.}

\end{figure}

\subsection{Comparisons and sensitivity of the results}

Here, we concentrate on analysing the sensitivity of the results presented in the previous section and on comparisons with baselines.

Table \ref{table:algo_results} shows the intermediate score on each of the goals for each of the reward signals every 1 million timesteps. This allows us to evaluate the sensitivity of the presented solutions, as we can investigate what could have happened, had the algorithm made different choices along the way. The table also includes the scores on the said 1 million checkpoints for three baseline DQN algorithms ('b100', 'b010', 'b001' in Table \ref{table:algo_results}) and a random play baseline (rand in Table \ref{table:algo_results}). The random play baseline takes random actions at each timestep. With equal probability it can move up, down, or stay it the same place. The baseline DQN algorithms have the same architecture as the DQN used in Algorithm \ref{alg:re}, however the reward signals on which they learn differ. The rewards for the baseline DQNs are based on the score. This means that 'b100' receives a reward of 1 only when it scores a point, 'b001' when it loses a point, and 'b010' receives a reward of 1 as long as neither of the players scores. The baseline algorithms 'b100', 'b010', 'b001' are closest - it would seem - to the semi-evolved algorithms that used '100', '010', and '001' as their respective reward signals. The main difference is in the frequency of the timeframes for which the algorithm receives the reward.

The results of the random play, as shown in Table \ref{table:algo_results} are rather consistent as far as all the three proposed goal functions are concerned.
To analyze the results in Table \ref{table:algo_results} in greater detail, we look at each of the goals separately.


\begin{table}[t]
\centering
\caption{Performance of DQN on the goal of winning, losing and cooperating for all evolved signal combinations, baseline score-based signals, and random signal at each 1M checkpoint between 2M and 9M time steps, when no signal elimination is performed. Additionally, average values of the score obtained per each goal across the whole training session are provided.}
\label{table:algo_results}

\scalebox{0.73}{
\begin{tabular}{lrrrrrrrrr}
\toprule
     & \multicolumn{1}{l}{}    & \multicolumn{1}{l}{2M}   & \multicolumn{1}{l}{}      & \multicolumn{1}{l}{}    & \multicolumn{1}{l}{3M}   & \multicolumn{1}{l}{}      & \multicolumn{1}{l}{}    & \multicolumn{1}{l}{4M}   & \multicolumn{1}{l}{}      \\
$r_j$ & \multicolumn{1}{l}{won} & \multicolumn{1}{l}{lost} & \multicolumn{1}{l}{co-op} & \multicolumn{1}{l}{won} & \multicolumn{1}{l}{lost} & \multicolumn{1}{l}{co-op} & \multicolumn{1}{l}{won} & \multicolumn{1}{l}{lost} & \multicolumn{1}{l}{co-op} \\ \midrule
000  & 0                       & 1366                     & 73.17                     & 0                       & 1375                     & 72.66                     & 0                       & 1372                     & 72.84                     \\
001  & 27                      & 1328                     & 73.78                     & 0                       & 1376                     & 72.47                     & 0                       & 1359                     & 73.54                     \\
010  & 79                      & 624                      & 142.06                    & 79                      & 541                      & 160.97                    & 68                      & 624                      & 144.33                    \\
100  & 312                     & 586                      & 111.05                    & 141                     & 523                      & 150.04                    & 284                     & 553                      & 119.26                    \\
011  & 0                       & 1123                     & 88.90                     & 2                       & 1235                     & 80.72                     & 0                       & 1369                     & 73.01                     \\
101  & 0                       & 1380                     & 72.42                     & 0                       & 1311                     & 76.21                     & 0                       & 1380                     & 72.42                     \\
110  & 137                     & 579                      & 139.36                    & 138                     & 416                      & 180.09                    & 66                      & 504                      & 175.07                    \\
111  & 0                       & 1275                     & 78.41                     & 2                       & 1369                     & 72.90                     & 1                       & 1385                     & 72.11                     \\
b100 & 62                      & 773                      & 119.54                    & 0                       & 1311                     & 76.24                     & 256                     & 895                      & 86.64                     \\
b010 & 152                     & 855                      & 99.18                     & 244                     & 344                      & 169.92                    & 110                     & 644                      & 132.37                    \\
b001 & 0                       & 1378                     & 72.50                     & 0                       & 1378                     & 72.47                     & 0                       & 1392                     & 71.92                     \\
rand & 62                      & 997                      & 94.24                     & 75                      & 965                      & 96.07                     & 62                      & 1006                     & 93.46                     \\ \midrule

  &                        &                      &                     &                        &                    &                      &                      &                     &                     \\

     & \multicolumn{1}{l}{}    & \multicolumn{1}{l}{5M}   & \multicolumn{1}{l}{}      & \multicolumn{1}{l}{}    & \multicolumn{1}{l}{6M}   & \multicolumn{1}{l}{}      & \multicolumn{1}{l}{}    & \multicolumn{1}{l}{7M}   & \multicolumn{1}{l}{}      \\
$r_j$ & \multicolumn{1}{l}{won} & \multicolumn{1}{l}{lost} & \multicolumn{1}{l}{co-op} & \multicolumn{1}{l}{won} & \multicolumn{1}{l}{lost} & \multicolumn{1}{l}{co-op} & \multicolumn{1}{l}{won} & \multicolumn{1}{l}{lost} & \multicolumn{1}{l}{co-op} \\ \midrule
000  & 3                       & 1120                     & 88.84                     & 0                       & 1373                     & 72.68                     & 0                       & 1372                     & 72.79                     \\
001  & 1                       & 1383                     & 72.14                     & 0                       & 1386                     & 72.10                     & 51                      & 1202                     & 79.72                     \\
010  & 163                     & 277                      & 226.97                    & 54                      & 481                      & 186.40                    & 99                      & 339                      & 227.83                    \\
100  & 341                     & 552                      & 111.79                    & 223                     & 431                      & 152.05                    & 394                     & 434                      & 120.65                    \\
011  & 13                      & 743                      & 132.07                    & 15                      & 1302                     & 75.81                     & 0                       & 1310                     & 76.23                     \\
101  & 0                       & 1359                     & 73.51                     & 0                       & 1370                     & 72.92                     & 0                       & 1369                     & 72.90                     \\
110  & 343                     & 272                      & 161.76                    & 192                     & 405                      & 167.16                    & 206                     & 294                      & 199.16                    \\
111  & 0                       & 1383                     & 72.19                     & 41                      & 1175                     & 82.08                     & 0                       & 1231                     & 81.20                     \\
b100 & 433                     & 304                      & 135.44                    & 550                     & 355                      & 110.37                    & 577                     & 267                      & 118.3                     \\
b010 & 262                     & 412                      & 148.15                    & 64                      & 547                      & 163.49                    & 102                     & 509                      & 163.11                    \\
b001 & 0                       & 1376                     & 72.57                     & 0                       & 1373                     & 72.78                     & 0                       & 1384                     & 72.21                     \\
rand & 72                      & 990                      & 94.06                     & 49                      & 986                      & 96.54                     & 68                      & 933                      & 99.84                     \\ \midrule

  &                        &                      &                     &                        &                    &                      &                      &                     &                     \\

     & \multicolumn{1}{l}{}    & \multicolumn{1}{l}{8M}   & \multicolumn{1}{l}{}      & \multicolumn{1}{l}{}    & \multicolumn{1}{l}{9M}   & \multicolumn{1}{l}{}      & \multicolumn{1}{l}{}    & \multicolumn{1}{l}{Avg.} & \multicolumn{1}{l}{}      \\
$r_j$ & \multicolumn{1}{l}{won} & \multicolumn{1}{l}{lost} & \multicolumn{1}{l}{co-op} & \multicolumn{1}{l}{won} & \multicolumn{1}{l}{lost} & \multicolumn{1}{l}{co-op} & \multicolumn{1}{l}{won} & \multicolumn{1}{l}{lost} & \multicolumn{1}{l}{co-op} \\ \midrule
000  & 0                       & 1382                     & 72.27                     & 0                       & 1374                     & 72.70                     & 0                       & 1342                     & 74.74                     \\
001  & 0                       & 1375                     & 72.62                     & 0                       & 1385                     & 72.14                     & 10                      & 1349                     & 73.56                     \\
010  & 48                      & 503                      & 180.99                    & 98                      & 338                      & 228.74                    & 86                      & 466                      & 187.29                    \\
100  & 494                     & 499                      & 100.44                    & 505                     & 385                      & 112.05                    & 337                     & 495                      & 122.17                    \\
011  & 2                       & 1381                     & 72.23                     & 0                       & 1382                     & 72.23                     & 4                       & 1231                     & 83.90                     \\
101  & 0                       & 1380                     & 72.40                     & 0                       & 1382                     & 72.33                     & 0                       & 1366                     & 73.14                     \\
110  & 194                     & 200                      & 253.17                    & 161                     & 286                      & 222.98                    & 180                     & 370                      & 187.34                    \\
111  & 0                       & 1370                     & 72.94                     & 0                       & 1381                     & 72.38                     & 6                       & 1321                     & 75.53                     \\
b100 & 528                     & 289                      & 122.09                    & 482                     & 289                      & 129.45                    & 361                     & 560                      & 112.26                    \\
b010 & 140                     & 331                      & 211.9                     & 83                      & 559                      & 155.55                    & 145                     & 525                      & 155.46                    \\
b001 & 0                       & 1392                     & 71.76                     & 0                       & 1382                     & 72.31                     & 0                       & 1382                     & 72.32                     \\
rand & 58                      & 999                      & 94.49                     & 65                      & 975                      & 96.07                     & 64                      & 981                      & 95.60                     \\ \bottomrule
\end{tabular}
}
\end{table}

\subsubsection{Learning to lose}
Losing every single point seems to be the easiest goal to learn, as expected. The reward signal '001' led to a very good result on the goal of losing the maximum number of points, but its performance was close on that goal to many other reward signals that just proved too difficult to learn, and similar to the baseline score 'b001'. Semi-evolved rewards '000' and '111' did not lead to any significant learning, even over 9 million timesteps of training - and as a result achieved a good performance on the goal of losing. The final results on all the goals for the said reward signals where close to that of '011' and '101', which also proved too confusing for the DQN to learn based on the same pixel input.

\subsubsection{Learning to win}
Three out of the eight semi-evolved reward signals led to an algorithm that was able to score more points than random play. Namely, '010', '100', '110'. It seems that as long as the reward is given for keeping the ball between the paddles, and there is no incentive to lose - the signal is good enough to learn to score some points. Interestingly, the signal '110' led to the second best performing algorithm on the goal concerning scoring as many points as possible. It was only topped by perhaps the ``obvious" reward signal choice of '100'. The baseline algorithm 'b100' proved better on average than '100' across the 8 test checkpoints, however it did suffer from less consistency, and even got stuck in a local extremum at 3 million timeframes (the algorithm kept the paddle in the lower part of the screen and only occasionally managed to hit the ball). This could be explained by the frequency of the timeframes for which the algorithm receives the reward during training. It seems that the performance of the algorithm trained on the '100' signal was more consistently improving on its goal, and achieved the highest score at 9 million timesteps, with the highest score on $G_1$ of all the algorithms. It seems it was also learning faster than the baseline in the beginning.

\subsubsection{Learning to cooperate}
As far as cooperation is concerned, two reward signals led to a very similar average performance on that goal. Concretely, running an algorithm to learn on the reward signal '010' and '110' would give comparable performance. We find it interesting, as we would argue that '110' is not an obvious choice to set-up a priori with cooperation in mind. Both reward signals led to a better cooperation than the baseline 'b010'. It is worth noting, that not only did the reward signal '110' lead to the highest average cooperation, but also to the third highest score on the ``winning" goal (including the baseline 'b100').

\subsubsection{Sensitivity}
Finally, Algorithm \ref{alg:re} led to the following choice of reward signals: '100', '000', '110' for $G_1$, $G_2$ and $G_3$, respectively. The results are not very sensitive to the choices between different reward signals along the way. This is because in the final round we always keep the solution leading to the best performance in that particular round, and in previous iterations we only eliminate the worst performing solution. This means that as long as the signal does not lead to the worst performance on a given goal consistently, it has a high probability of making it through to the next round. The signals '100', '000', '110' lead to a very good performance on their respective goals in each round and are the best in the final round, as a result they emerge as the winners of the algorithm unfailingly. Those signals can be seen as dominant strategies for their particular goals and they end up surviving no matter at which stage they were introduced. In comparison, due to its low performance at one of the milestones, the baseline signal 'b100' does not progress to the final elimination round 39\% of the time.

For clarity, we summarize the main results of the experiment in the list below:
\begin{itemize}
\item Algorithm \ref{alg:re} proved to work on highly dimensional input space of pixels
\item many rewards can lead to a good performance on the goal of ``losing"
\item fewer reward signals lead to learning to cooperate
\item the goal of ``winning" proved the most difficult to learn, with the smallest number of reward signals leading to the desired outcome
\item our experiments give further evidence of the fact that there exist multiple reward signals to learn a given goal
\item interestingly, the performance during training of the baseline 'b100' seemed much less stable and predictable than for the algorithm trained on the reward signal '100', possibly due to the frequency of timesteps for which each algorithm ``experienced" the reward
\item lastly, the sensitivity of the results shows that it is not obvious at which point the training phase should stop, in order to have the agent performing at its best with respect to the goal.
\end{itemize}

\section{Discussion}

The problem of reward signal design is an interesting open area of research in RL, especially as we move from low towards highly dimensional input spaces where it is potentially easier to misspecify the reward and end up with an unwanted behaviour of the algorithm.

As a possible approach for exploring the space of all possible reward signals, in this work we presented an algorithm that allows to mutate the reward at each timestep of interacting with the environment. We showed that it is possible to use the proposed algorithm in order to learn complex behaviours of winning, cooperation and losing by specifying high-level goal functions, but without providing a concrete reward signal for any of them. The reward signals were discovered from scratch and we found that often there is no unique reward signal to achieve a certain level of performance on a given goal. The number of reward signals leading to a given goal also varies, depending on the complexity of the goal.

The proposed algorithm is highly adaptable, i.e. the designer can change the learning algorithm used, modify the goals (also look at combinations of basic goals for a specific application), change the statistic used as the fitness function, control the number of mutations, etc.

There are many possible extensions of the presented approach. First, sampling the reward from a continuous rather than discrete set of values seems worth investigating as are any ways for lowering the computational cost of the algorithm. Additionally, it has been observed that training using independent RL can lead to overfitting to opponents' strategy \cite{lanctot2017unified,muszynski2017happiness} and one way to overcome that problem is to use population-based training \cite{jaderberg2018human}. Hence, we argue that joining reward evolution with population-based training could improve such solutions to an even
greater degree.

\section*{Acknowledgments}
This work was supported by Microsoft Research through its PhD Scholarship Programme.

\bibliographystyle{named}
\bibliography{ijcai21}

\end{document}